\documentstyle[12pt]{article}
\begin {document}
\title{Boundary K-matrices for the six vertex and the $n(2n-1)$
 $A_{n-1}$ vertex models}
\author{H.J. de Vega * \and A. Gonz\'alez Ruiz **}
\date{LPTHE-PAR 92-45, November 1992}

\maketitle
\begin {abstract}
Boundary conditions compatible with integrability are obtained for
two dimensional models by solving the factorizability equations
for the reflection matrices $K^{\pm}(\theta)$. For the six vertex
model the general solution depending on four arbitrary parameters
is found. For the $A_{n-1}$ models all diagonal
solutions are found. The associated integrable magnetic Hamiltonians
are explicitly derived.
\end{abstract}

\vspace{5cm}
telnet parthe.lpthe.jussieu
* L.P.T.H.E , Tour 16, 1er. \'etage, Universit\'e Paris VI,
4, Place Jussieu, 75252 Paris cedex 05, FRANCE and
Isaac Newton Institute, 20 Clarkson Road, Cambridge, CB3 OEH, U. K.

** Departamento de F\'isica Te\'orica,
Universidad Complutense, 28040 Madrid , SPAIN.

\pagebreak

\section{Introduction}
As is by now well known integrability is a consequence of the Yang-Baxter
equation in two-dimensional lattice models and quantum field theory \cite{h}.
The Yang-Baxter equation takes the form:

\begin{equation}
\begin{array}{l}
[1\otimes R(\theta-\theta^\prime)][R(\theta)\otimes 1][1\otimes
R(\theta^\prime)]=\\
{[}R(\theta^\prime)\otimes 1][1\otimes R(\theta)]
[R(\theta-\theta^\prime)\otimes 1]  \label{yb}
\end{array}
\end{equation}

where the R matrix elements $R^{ab}_{cd}(\theta)$ with
$(1\leq a,b,c,d\leq n, n\geq 2)$ define the statistical
weights for a vertex model in two dimensions.\\ Not all boundary
conditions (bc) are compatible with integrability in the bulk.
 Eq. (\ref{yb}) guarantees the integrability on the
bulk. Integrability
holds for bc defined by matrices $K^{-}(\theta)$ and $K^{+}(\theta)$
(associated to the left and right boundaries respectively) provided
the R matrix has P,T and crossing symmetry and the following equations
proposed by Cherednik and Sklyanin \cite{i} are fullfilled:

\begin{equation}
\begin{array}{l}
R(\theta-\theta^\prime)[K^{-}(\theta)\otimes
1]R(\theta+\theta^\prime)[K^{-}(\theta^\prime)\otimes 1]=\\
{[}K^{-}(\theta^\prime)\otimes 1]R(\theta+\theta^\prime)[K^{-}(\theta)\otimes
1]R(\theta-\theta^\prime) \label{sk1}
\end{array}
\end{equation}

\begin{equation}
\begin{array}{l}
R(\theta-\theta^\prime)[1\otimes K^{+}(\theta) ]R(\theta+\theta^\prime)[
1\otimes K^{+}(\theta^\prime)]=\\ {[} 1\otimes K^{+}(\theta^\prime)]
R(\theta+\theta^\prime)[1\otimes K^{+}(\theta) ]R(\theta-\theta^\prime)
\label{sk2}
\end{array}
\end{equation}

In addition the Yang-Baxter equation (\ref{yb}) guarantees the factorizability
 for the S matrix where:

\begin{equation}
S^{ab}_{cd}(\theta)=R^{ab}_{dc}(\theta)
\end{equation}

is a two-particle S-matrix in two spacetime dimensions. In this context,
$K^-_{ab}(\theta)$ and $K^+_{ab}(\theta)$ describe the scattering of the
particles by the left and right boundaries respectively. Therefore for each
integrable model (a given solution to the Yang-Baxter equation $R(\theta)$) in
order to find the integrable boundary conditions , one must find the solutions
$K^-_{ab}(\theta)$ and $K^+_{ab}(\theta)$ of eqs. (\ref{sk1}) and (\ref{sk2})
for the given $R(\theta)$.\\
 We present in this note the general solution of
eqs.(\ref{sk1})-(\ref{sk2}) for the six vertex model and a
family of solutions to the corresponding modifications of these
equations for the $A_{n-1}$ vertex model \cite{h}. Also the associated
magnetic hamiltonians are derived and their properties investigated.

\section{General $K^{\pm}$ matrices for the six vertex model and its associated
hamiltonians}

Let us first consider the six vertex model \cite{h}. The R matrix has the form:

\begin{equation}
R(\theta)=\left(\begin{array}{cccc}
1 & 0 & 0 & 0\\
0 & \frac{\sinh{\gamma}}{\sinh{(\theta+\gamma)}} &
\frac{\sinh{\theta}}{\sinh{(\theta+\gamma)}} & 0\\
 0 & \frac{\sinh{\theta}}  {\sinh{(\theta+\gamma)}} &
\frac{\sinh{\gamma}}  {\sinh{(\theta+\gamma)}}  & 0\\
0 & 0 & 0 & 1
\end{array}\right) \label{R}
\end{equation}

Since this R matrix enjoys P symmetry :

\begin{equation}
PR(\theta)P=R(\theta)
\end{equation}

eq.(\ref{sk1}) and eq.(\ref{sk2}) are equivalent for the six vertex model.\\
We seek for the general solution of these equations:

\begin{equation}
K(\theta)=\left(\begin{array}{cc}
x(\theta) & y(\theta)\\
z(\theta) & t(\theta)
\end{array}\right) \label{K}
\end{equation}

where $x(\theta)$, $y(\theta)$, $z(\theta)$ and $t(\theta)$ are unknown
functions.
Inserting eqs. (\ref{R}) and (\ref{K}) in eqs.(\ref{sk1}) or (\ref{sk2})
yields ten functional equations for these four functions.\\
The relevant ones are:

\begin{equation}
z(\theta)y(\theta^\prime)=z(\theta^\prime)y(\theta)  \label{c1}
\end{equation}

\begin{equation}
\sinh(\theta-\theta^\prime)[x(\theta)x(\theta^\prime)-t(\theta)t(\theta^\prime)]+
\sinh(\theta+\theta^\prime)[x(\theta^\prime)t(\theta)-x(\theta)t(\theta^\prime)]=0
\label{c2}
\end{equation}

\begin{equation}
y(\theta)x(\theta^\prime)\sinh2\theta^\prime=
[\sinh{(\theta+\theta^\prime)} x(\theta) +
\sinh{(\theta-\theta^\prime)} t(\theta)] y(\theta^\prime)
\label{c3}
\end{equation}

Eq.(\ref{c1}) imply:

\begin{equation}
y(\theta)=k_{1}z(\theta)
\end{equation}

where $k_{1}$ is an arbitrary constant.\\
Eq.(\ref{c2}) can be rewritten as:

\begin{equation}
[\tanh(\theta)-\tanh(\theta^\prime)][1-a(\theta)a(\theta^\prime)]+
[\tanh(\theta)+\tanh(\theta^\prime)][a(\theta)-a(\theta^\prime)]=0 \label{c2p}
\end{equation}

where $a(\theta)=t(\theta)/x(\theta)$. Differentiating eq.(\ref{c2p})
 with respect
to $\theta^\prime$ and setting $\theta^\prime=0$ yields an algebraic equation
with solution:

\begin{equation}
\frac{t(\theta)}{x(\theta)}=\frac{\sinh(\xi-\theta)}{\sinh(\xi+\theta)}
\end{equation}

where $\xi$ is another arbitrary constant.
Then, eq.(\ref{c3}) tells us that:

\begin{equation}
y(\theta)=\mu \sinh2\theta
\end{equation}

The remaining equations are identically satisfied. In summary,
 the general solution $K(\theta)$ for the six-vertex model can be
written as:

\begin{equation}
K(\theta,k,\lambda,\mu,\xi)=\left( \begin{array}{cc}
k \sinh(\xi+\theta) & \mu \sinh2\theta \\
\lambda \sinh2\theta & k \sinh(\xi - \theta)
\end{array}  \right) \label{KG}
\end{equation}

where $k$,$\lambda$,$\mu$ and $\xi$ are arbitrary parameters.\\
The special case $\lambda=\mu=0$ reproduces the solution given
 in ref.\cite{i}.\\

Let us now discuss the integrable hamiltonians associated to the
K-matrix (\ref{KG}). They follow by the procedure discused in
\cite{i} from the equation:

\begin{equation}
H=C\left\{\sum^{N-1}_{n=1}h_{n,n+1}+\frac{1}{2} \dot{K}^{-}_{1}(0)+
\frac{tr_{0}[K^{+t}_{0}(-\eta)h_{N0}]}{tr[K^{+}(-\eta)]}\right\}
\label{H} \end{equation}

Here C is an arbitrary constant and :
\begin{equation}
h^{n,n+1}=\dot{R}_{n,n+1}(0)
\end{equation}

gives the two sites hamiltonian in the bulk and the indices $(n,n+1)$
label the sites in wich the matrix acts. In the present, case we can
choose:

\begin{equation}
K^{\pm}(\theta)=K(\theta,k_{\pm},\lambda_{\pm},\mu_{\pm},\xi_{\pm}) \label{Kpm}
\end{equation}

If we want to maintain the bulk part of the XXZ hamiltonian with
the first derivative of the transfer matrix we are led to take
 $K^{-}$ matrices with value in $\theta=0$ diferent from zero,
this leads us to $k_{-}\neq 0$ and without loss of generality we
can take $K^{-} (0)=1$ that is $k_{-}= 1/{\sinh\xi}$. For the same
reason we choose $k_{+}\neq 0$.

Inserting eqs. (\ref{R}),(\ref{KG}) and (\ref{Kpm}) in eq. (\ref{H}) yields:

\begin{eqnarray}
H =
\sum_{n=1}^{N-1}\left(\sigma^{x}_{n}\sigma^{x}_{n+1}+
\sigma^{y}_{n}\sigma^{y}_{n+1}+
\cosh\gamma \; \sigma^{z}_{n}\sigma^{z}_{n+1}\right)
\nonumber \\
+\sinh\gamma\left(b_{-}\sigma_{1}^{z}-b_{+}\sigma_{N}^{z}+
c_{-}\sigma_{1}^{-}-c_{+}\sigma_{N}^{-}+
d_{-}\sigma_{1}^{+}-d_{+}\sigma_{N}^{+}\right)
\label{Hpm}
\end{eqnarray}

we have chosen $C=2 \sinh\gamma$ and omitted the terms proportional to the
identity. The parameters $b_{\pm}$, $c_{\pm}$ and $d_{\pm}$ follow
from $\lambda_{\pm}$, $\mu_{\pm}$, $\xi_{\pm}$ and $k_{+}$ as shown:

\begin{eqnarray}
b_{-}=\coth\xi_{-}\nonumber\\
b_{+}=\coth\xi_{+}\nonumber\\
c_{-}=2\lambda_{-}\nonumber\\
c_{+}=\frac{2\lambda_{+}}{k_{+}\sinh\xi_{+}}\nonumber\\
d_{-}=2\mu_{-}\nonumber\\
d_{+}=\frac{2\mu_{+}}{k_{+}\sinh\xi_{+}}
\end{eqnarray}

The bulk part in eq.(\ref{Hpm}) is just the well-known XXZ Heisenberg
hamiltonian.
When $c_{\pm}=d_{\pm}=0$ we recover the hamiltonian discussed in
ref \cite{i}.
Equation (\ref{Hpm}) provides the more general choice of boundary
terms compatible with integrability for the XXZ chain besides periodic
and twisted bc. It contains
three  parameters in each boundary ($b_{\pm}$,$c_{\pm}$ and
$d_{\pm}$). In particular when $b_{\pm}=\pm 1$ and
$c_{\pm}=d_{\pm}=0$, one recovers the $SU_{q}(2)$ invariant
hamiltonian ($q=e^{\mp\gamma}$) \cite{v}, \cite{c}.

\section{ Diagonal $K^{\pm}(\theta)$ matrices for the $A_{n-1}$
vertex models}

Let us now consider the $n(2n-1)$ integrable vertex model asociated
to the $A_{n-1}$ Lie algebra. The $R$ -matrix takes the form
(see ref.\cite{h})

\begin{eqnarray}
R^{ab}_{ij}(\theta) = \frac{\sinh \gamma}{\sinh( \gamma + \theta)} \
\delta_{ia} \delta_{jb} e^{ \theta sign (a-b)}  +
\frac{\sinh \theta}{\sinh( \gamma +\theta)}  \ \delta_{ib}
\delta_{ja}\quad , \quad a\neq b \nonumber\\
R^{aa}_{ii} = \delta_{ia}   \hspace{1cm}
1\leq a,b \leq n \label{RN} \end{eqnarray}

This reduces to eq.(\ref{R}) when $n = 2$ upon a gauge
transformation \cite{h}. Contrary to the six vertex $R$-matrix, the
$R$ -matrix (\ref{RN}) does not enjoy $P$ and $T$ symmetry but just $PT$
invariance. It is not crossing invariant either but it obeys the
weaker property \cite{n}.

\begin{equation}
\left[\left\{\left[S_{12}(\theta)^{t_{2}}\right]^{-1}\right\}^
{t_{2}}\right]^{-1}
=
L(\theta,\gamma) M_{2} S_{12}(\theta +2\eta) M_{2}^{-1}
\label{SM}
\end{equation}

where $L(\theta,\gamma)$ is a c-number function, $\eta$ a constant and $M$ a
symmetry of the $R$ -matrix (\ref{RN}). That is

\begin{equation}
[M_{1} \otimes M_{2}, R_{12}(\theta)] = 0
\label{sy}
\end{equation}

We find by direct calculation from eqs.(\ref{RN}) and (\ref{SM}):

\begin{eqnarray}
\eta & = & \frac{n}{2} \gamma\nonumber\\
M_{ab} & = & \delta_{ab} \  e^{(n-2a+1)\gamma} \hspace{2cm} 1 \leq a,b \leq n
\nonumber \\
L(\theta,\gamma) & = & \frac{\sinh(\theta+\gamma) \sinh[\theta + (n-1)
\gamma]}{ \sinh(\theta) \sinh(\theta + n \gamma)} \label{M} \nonumber \\
\end{eqnarray}

In this case, when the weak condition (\ref{SM}) holds, the
integrability requirements on the boundary matrices $K^{-}(\theta)$
and $K^{+}(\theta)$ need some modifications \cite{l}. $K^{-}(\theta)$
must still obey eq (2) and $K^{+}(\theta)$ obeys:

\begin{equation}
\begin{array}{l}
R(\theta-\theta^{\prime})K^{+}_{1}(\theta^{\prime})^{t_{1}}
M_{1}^{-1} R(-\theta -\theta^{\prime} -2\eta)
K^{+}_{1}(\theta)^{t_{1}}M_{2}=\\
 K^{+}_{1}(\theta)^{t_{1}} M_{2} R(- \theta -\theta^{\prime} -2\eta)
M_{1}^{-1} K_{1}^{+}(\theta^{\prime})^{t_{1}}  R(- \theta
-\theta^{\prime})\label{RM}
\end{array}
 \end{equation}

There is an automorphism beteewn $K^{-}$ and $K^{+}$

\begin{equation}
K^{+}(\theta) = K^{-}(- \theta -\eta)^{t} M\label{IS}
\end{equation}

For simplicity, we start for searching diagonal matrices  $K^{-}$ :

\begin{equation}
K^{-}_{ab}(\theta) = \delta_{ab} K^{-}_{a}\label{KA}
\end{equation}

Inserting eqs. (\ref{RN}) and (\ref{KA}) in (\ref{sk1}) yields:

\begin{equation}
\begin{array}{l}
\sinh(\theta+\theta^{\prime}) [K_{a}^{-}(\theta)
K_{b}^{-}(\theta^\prime) e^{sign(a-b)(\theta-\theta^{\prime})}  -
K_{a}^{-}(\theta^{\prime}) K_{b}^{-}(\theta)
e^{-sign(a-b)(\theta-\theta^{\prime} ) }] +\\
\sinh(\theta-\theta^{\prime}) [K_{b}^{-}(\theta)
K_{b}^{-}(\theta)^{\prime} e^{-sign(a-b)(\theta+\theta^{\prime})} -
K_{a}^{-}(\theta^\prime) K_{a}^{-}(\theta)
e^{sign(a-b)(\theta^{\prime} + \theta) }]  =  0 \label{eq}
\end{array}
\end{equation}

These equations  are
generalizations of eq. (\ref{c2}). By a similar procedure we find
their general solution:

\begin{eqnarray}
K^{-}_{a}(\theta) = k \ \sinh(\xi + \theta) \ e^{\theta}  \hspace{2cm}
 1\leq a\leq l_{-} \nonumber\\
K^{-}_{a}(\theta) =  k \ \sinh(\xi - \theta) \ e^{-\theta}  \hspace{2cm}
 l_{-}+1\leq a\leq n \label{KMN}
\end{eqnarray}

Here $k$, $\xi$ and $l_{-}$ are arbitrary parameters. For $n=2$ and
$l_-=1$ we recover the diagonal case of eq.(\ref{KG}) after a
gauge transformation. In general, for $n > 2$, we have the extra
discrete parameter $l_{-}$ that tells where the diagonal element change
from one to another.

The integrable hamiltonian associated to the $R$ -matrix (\ref{RN})
with b.c. (\ref{KMN}) follows using equation (\ref{H}).
We find after some straightforward calculations:

\begin{eqnarray}
H = \sum_{j=1}^{N-1} \left\{ \cosh\gamma\sum_{r=1}^{n} \ e_{rr}^{(j)} \
e_{rr}^{(j+1)} + \sum_{\begin{array}{c} \scriptsize
r,s=1 \\ \scriptsize r \neq s \end{array}}^{n}  e_{rs}^{(j)} \
e_{sr}^{(j+1)} +  \sinh \gamma \sum_{ r,s=1}^{n} \ sign(r-s)
e_{rr}^{(j)}  e_{ss}^{(j+1)} \right\}         \nonumber \\ +
\frac{\sinh\gamma}{2} (1 + \coth \xi_{-}) \left[\sum_{r=1}^{l_{-}}
e_{rr}^{(1)} -  \sum_{l_{-}+1}^{n} e_{rr}^{(1)}\right] +
\frac{\cosh\gamma}{trK^{+}(0)}
 \{ \frac{\sinh(\xi_{+} - \frac{n\gamma}{2})}{\sinh\xi_{+}} e^{\gamma}
\sum_{r=1}^{l_{+}}  e_{rr}^{(N)} e^{-2r \gamma}
\nonumber \\
+ \frac{\sinh(\xi_{+} + \frac{n\gamma}{2})}{\sinh\xi_{+}} e^{(n+1)\gamma}
\sum_{r=l_{+}-1}^{n}  e_{rr}^{(N)} e^{-2r \gamma} \}  +
\frac{1}{trK^{+} (0)}   \sum_{ r,s=1}^{n} \ sign(r-s) e_{rr}^{(N)} \
K_{s}^{+}(0)   \nonumber \\ \end{eqnarray}

Where $N$ is the number of sites of the chain and $l_{+}$, $l_{-}$
arbitrary integers running from 1 to $n$. We have extracted a
global factor of $1/{\sinh\gamma}$ and omitted the terms proportional to the
unit operator.\\
 In particular with
$\xi_{\pm} =- \infty$ we get a $SU_{q}(n)$ invariant hamiltonian,
with $q=e^{- \gamma}$, that can be written as:

\begin{eqnarray}
H = \sum_{j=1}^{N-1} \{ \sum_{\begin{array}{c} \scriptsize r,s=1
\\
 \scriptsize r > s \end{array}}^{n}  ( \prod_{l=s}^{r-1} \
(J_{l}^{+})^{(j)} \prod_{l=r-1}^{s} (J_{l}^{-})^{(j+1)} +
\prod_{l=r-1}^{s} (J_{l}^{-})^{(j)} \prod_{l=s}^{r-1} \
(J_{l}^{+})^{(j+1)}) + \nonumber \\
 \frac{\cosh\gamma}{n} [ \sum_{ \begin{array}{c} \scriptsize r,s=1
\\ \scriptsize r > s \end{array}}^{n-1} s(n-r)
(h_{r}^{(j)}h_{s}^{(j+1)} + h_{s}^{(j)}h_{r}^{(j+1)}) +
\sum_{r=1}^{n-1} r(n-r) h_{r}^{(j)}h_{r}^{(j+1)} ] + \nonumber
\\
  \frac{\sinh\gamma}{n}\sum_{\begin{array}{c} \scriptsize r,s=1
\\
 \scriptsize r > s \end{array}}^{n-1} s(r-s)(n-r)(
h_{r}^{(j)}h_{s}^{(j+1)} - h_{s}^{(j)}h_{r}^{(j+1)}) \} +
\frac{\sinh\gamma}{n}\sum_{r=1}^{n-1} r(n-r) (
h_{r}^{(N)}-h_{r}^{(1)} )  \nonumber \\
\end{eqnarray}

Here $N$ is the number of sites, $J_{l}^{+} \equiv  e_{l
l+1}$, $J_{l}^{-} \equiv e_{l+1 l}$ and $h_{l} \equiv e_{l l}-e_{l+1 l+1}$
are the
$su(n)$ generators in the fundamental representation with $(e_{l
m})_{i j} \equiv  \delta_{l i}\delta_{m j}$. It is easily seen that this
hamiltonian coincides for $n=2$ with the $SU_{q}(2)$ invariant one, discussed
in
\cite{v},\cite{c}.

\section{Conclusions}
We have obtained the general solution to the surface factorization equations
for the six-vertex R matrix providing in this way the more general
boundary terms compatible with integrability. The Bethe ansatz in these systems
must change drastically
as the hamiltonian does not commute with $J_{z}$. For the
$A_{n}$ chain, a generalization of the nested Bethe
Ansatz \cite{h} will be needed.
\vspace{3cm}

A.G.R. would like to thank the LPTHE for the kind hospitality and the Spanish
M.E.C for financial support under grant AP90 02620085.
H. J. de V. would like to thank the Isaac Newton Institute for their
kind hospitality.

Nota Added: After completion of this paper, we hear from A. B. Zamolodchikov
that he has independently obtained eq.(\ref{KG}) for the six-vertex model.

\end{document}